\documentclass[11pt,a4paper]{article}

\usepackage{graphicx,here}
\usepackage{amsmath,amssymb,latexsym,jcappub}
\usepackage{bm}

\newcommand{\nat}{Nature}
\newcommand{\apj}{ApJ}

\begin{document}



\title{Dark Matter And The Habitability of Planets}

\author[a,b]{Dan Hooper }
\author[a]{ and Jason H. Steffen}
\affiliation[a]{Center for Particle Astrophysics, Fermi National Accelerator Laboratory, Batavia, IL 60510}
\affiliation[b]{Department of Astronomy and Astrophysics, The University of Chicago, Chicago, IL  60637} 
\emailAdd{dhooper@fnal.gov}
\emailAdd{jsteffen@fnal.gov}

\abstract{
In many models, dark matter particles can elastically scatter with nuclei in planets, causing those particles to become gravitationally bound.  While the energy expected to be released through the subsequent annihilations of dark matter particles in the interior of the Earth is negligibly small (a few megawatts in the most optimistic models), larger planets that reside in regions with higher densities of slow moving dark matter could plausibly capture and annihilate dark matter at a rate high enough to maintain liquid water on their surfaces, even in the absence of additional energy from starlight or other sources.  On these rare planets, it may be dark matter rather than light from a host star that makes it possible for life to emerge, evolve, and survive.
}


\arxivnumber{1103.5086}

\maketitle

\section{Introduction}

Life, as generally imagined, is capable of evolving and surviving only under a limited range of environmental circumstances.  In particular, the presence of liquid water to serve as a universal solvent appears to be a likely requirement of carbon-based life.  This need, in turn, implies a finite range of temperatures under which it will be possible for life to emerge and survive.

The surface temperature of a typical planet is maintained primarily by light from a host star.  For planets with an Earth-like albedo and emissivity, and in orbit around a Sun-like star, one can calculate that average surface temperatures between 273 and 373 K will persist for orbital distances between approximately 0.6 and 1.1 AU.  This represents a naive estimate for the habitable zone of a Sun-like star---planets in significantly larger or smaller orbits will likely contain water in only solid or gaseous form.

Of course there are a number of caveats that can alter the boundaries of the habitable zone, such as chemical cycles~\citep{Kasting:1993} or the greenhouse effect.  For example, a more efficient greenhouse effect (leading to a lesser emissivity) than is found on the Earth could significantly increase a planet's surface temperature.  Venus, for instance, orbits at a distance of 0.72 AU and thus falls within the Sun's nominal habitable zone, but maintains an average surface temperature that is well above 700 K.  Alternatively, energy sources other than starlight could contribute to maintaining a planet's surface temperature.  The decay of radioactive elements and other sources of geothermal energy, for example, contribute approximately 0.025\% of the total energy that goes into maintaining the Earth's surface temperature~\citep{Pollack:1993}, but could potentially contribute more significantly for other planets.  A different atmospheric composition and thickness may provide a habitable environment even on rogue planets in the interstellar medium~\citep{Stevenson:1999}.  On some planets and moons, significant quantities of heat could also be generated by tidal flexing and other geological activity~\citep{Peale:1978,Murray:2000,Abbot:2011dz}.

In this paper, we consider another possible energy source for planetary heating---the annihilations of dark matter particles.  The mass of the dark matter contained in our universe represents an enormous energy reservoir---a factor of approximately $10^3$ times greater than the total energy that would be released through the fusion of all of the universe's hydrogen into helium.  But unlike baryonic matter, dark matter does not generally interact at rates sufficient to produce ecologically relevant quantities of energy.  An exception to this conclusion, however, could possibly be found for dark matter particles that have become gravitationally captured in a planet's interior.  Dark matter, in the form of weakly interacting massive particles (WIMPs), is generally predicted to interact with nuclei, enabling them to lose momentum and become gravitationally bound and captured by stars or planets~\citep{Silk:1985ax,Krauss:1985ks}.  After accumulating in a planet's interior, these dark matter particles can, in many models, subsequently annihilate to produce energetic particles that are then absorbed by the surrounding material. 

In the case of the Earth, dark matter annihilations are not expected to meaningfully compete with the total energy absorbed from sunlight~\citep{Mack:2007xj}.  Even for the most optimistic of the hypothesized dark matter candidates, we estimate that the Earth would gravitationally capture less than one billionth of the dark matter required to match the energy input from the Sun.  This conclusion, however, could be very different for planets in other galactic environments.  In particular, the density of dark matter is expected to be hundreds or thousands of times larger in the inner tens of parsecs of the Milky Way and in the cores of dwarf spheroidal galaxies than it is in our solar system.  This fact, combined with much lower velocities of dark matter particles in such environments (which helps to facilitate dark matter capture), could enable dark matter particles to accumulate in planetary objects at rates that would provide the dominant energy source for planets, potentially enabling their surfaces to maintain liquid water without the aid of light from a stellar host.  We find that rocky super-Earths in the inner region of the Milky Way could capture dark matter at a rate sufficient to add geologically meaningful heat to the planet, and in the cores of dwarf spheroidal galaxies the caputre rate may be sufficient to maintain liquid water on a planet's surface through the annihilations of dark matter alone.  Thus, rocky planets on large, cold orbits---whether scattered from the interior regions of a system or distant cores that failed to accrete a large envelope of light elements \citep{Laughlin:2004}---or rogue planets with no stellar host could potentially be habitable through this mechanism.  Moreover, the timescale over which such an object can maintain this surface temperature may exceed trillions of years---outliving even the smallest, and longest-lived, main sequence stars~\citep{Laughlin:1997}.  In this rather special and rare class of planets, the annihilations of dark matter could provide the energy necessary for liquid water, and thus life, to exist.

The remainder of this article is structured as follows. In Sec.~\ref{2} we discuss the capture of dark matter particles in Earth-like and super-Earth planets.  In Sec.~\ref{3}, we calculate the resulting surface temperature of planets in regions such as dwarf spheroidal galaxies and the inner Milky Way, finding that the surface temperatures can be significantly augmented in these regions, and allowing liquid water to plausibly be maintained on some planets, even in the absence of light from a host star.  In Sec.~\ref{discussion} we discuss some of the implications of these results and summarize our conclusions.

\section{The Capture Of Dark Matter Particles In Earth-like and Super-Earth Planets}
\label{2}

As dark matter particles move through the Galaxy, they can occasionally scatter elastically with nuclei within stars or planets.  In some of these collisions, enough momentum can be transferred as to leave the dark matter particle gravitationally bound to the object.  The subsequent orbits of such particles, which pass through the volume of the star or planet, lead to further scattering and eventually cause the dark matter to accumulate in the object's interior.  Once a sufficient quantity of dark matter is captured, it can begin to annihilate efficiently, converting its mass into kinetic energy in the form of relativistic particles that are then absorbed by the surrounding material.  To calculate the rate at which dark matter particles are expected to be captured by a planet, we follow the standard approach of \citet{capture1,capture2} (see also \cite{capture3}).


As we are most interested in those planets with chemistry, surface gravity, and other characteristics similar to the Earth (as such features are likely to be related to the likelihood of a planet being able to support life), we focus on broadly Earth-like planets.  Specifically, we consider rocky planets with an iron core, and with a range of masses between roughly 1 and 10 times that of the Earth. To describe the mass density of such planets, we adopt the models described in \citet{densitypro1,densitypro2}.  In particular, we use these models to set the overall radius, core radius, and average core density of a given planet. For simplicity, we assume a constant density for the core and mantle of each planet (this assumption leads to a modest underestimate of the capture rate of dark matter particles).  We adopt a core composition that is similar to that of the Earth: 89\% iron, 6\% nickel, and 5\% sulfur. The remaining volume of each planet is assumed to be made up of mostly oxygen (45\%), magnesium (23\%), silicon (22\%), iron (6\%), aluminium (2\%), and calcium (2\%).

The dark matter capture rate of a given planet depends on a number of environmental factors, as well as on the properties of the dark matter particles themselves.  In the case of the local galactic neighborhood, the average density of dark matter is inferred from galactic rotation curves to be approximately 0.4 GeV/cm$^3$---roughly $7\times 10^{-25}~$g/cm$^3$ or half of a proton mass per cubic centimeter---(see, for example, \citet{Catena:2009mf}). The velocity distribution can be approximated by a Maxwell-Boltzmann distribution with a velocity dispersion of roughly 250 km/s and shifted to account for the planet's motion relative to the frame of the dark matter halo.

For the properties of the dark matter particles themselves, we adopt two phenomenological models: a particle with a mass of 300 GeV and a spin-independent elastic scattering cross section with nucleons of $\sigma_0 = 8 \times 10^{-44}$ cm$^2$ (Model A), and a much lighter dark matter candidate with a mass of 7 GeV and a spin-independent elastic scattering cross section with nucleons of $\sigma_0 = 2 \times 10^{-40}$ cm$^2$ (Model B).  Model A can be taken as a somewhat typical example of a dark matter candidate that might appear in electroweak-scale extensions of the standard model of particle physics---supersymmetry, for example.  Model B is motivated by the signals reported by the CoGeNT~\citep{cogent} and DAMA/LIBRA~\citep{dama} collaborations~\citep{cogentdama}.  In each of these models, the dark matter particles interact with nuclei at approximately the maximal degree consistent with constraints from direct detection experiments~\citep{cdms,xenon,cdmslow}.  In the calculation of the differential scattering cross sections, we adopt form factors based on those described by \cite{formfactor} (although these factors are nearly negligible in our calculations).

For these dark matter models and the locally inferred distribution, we calculate that dark matter will be captured by the Earth at a rate given by approximately $3\times 10^{12}$ ($5\times 10^{15}$) particles per second in the case of dark matter model A (model B).  These rates are in good agreement with those calculated elsewhere (see \cite{Lundberg:2004dn}, for example).  While these numbers may seem large at first glance, they do not represent nearly enough energy to contribute significantly to the Earth's temperature.  Even if dark matter particles were to annihilate at the same rate at which they are captured, they would contribute a total input power of $1.4\times 10^{5}$ Watts ($5.6\times 10^6$~W).  In contrast, $1.74 \times 10^{17}$~W reach the Earth's atmosphere from the Sun (of which about $1.2 \times 10^{17}$~W is absorbed).  Furthermore, the dark matter annihilation rate will come into equilibrium with the capture rate only if the combined capture rate and annihilation cross sections are sufficiently large; in many models this equilibrium is not expected to be realized, further suppressing the power contributed through dark matter annihilation.

\begin{figure}[!ht]
\includegraphics[width=0.45\textwidth]{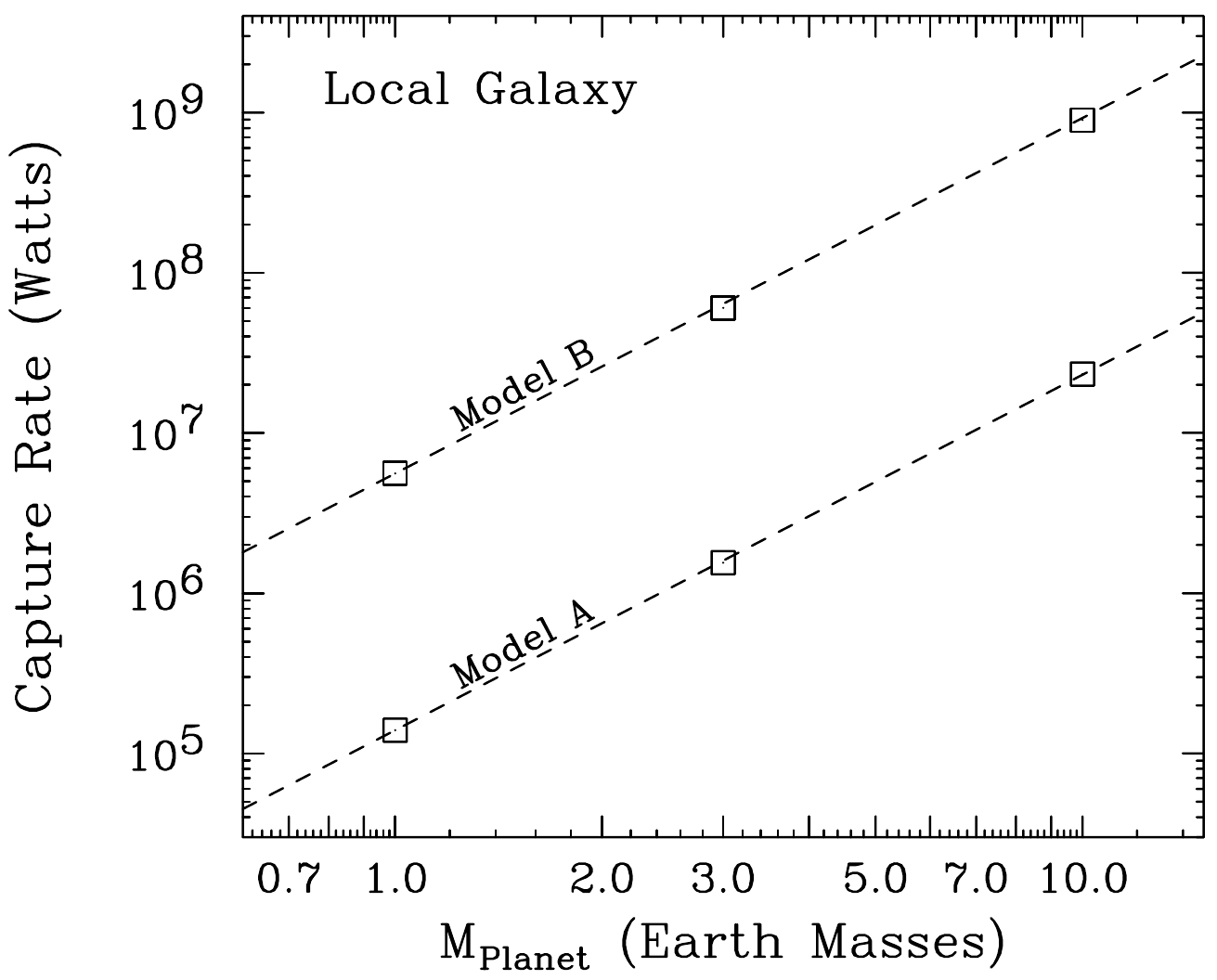}
\caption{The rate at which dark matter particles are predicted to be captured by an Earth-like planet in the local neighborhood of the Milky Way (in energy-equivalent units), as a function of the planet's mass. We show results for two optimistic dark matter models. In no case does the quantity of energy in dark matter captured remotely compete with the energy absorbed from the Earth by sunlight. See text for more details. \label{local}}
\end{figure}

The predicted capture rates of dark matter can be considerably higher, however, if one considers somewhat larger planets.  A ten Earth-mass planet in our galactic neighborhood, for example, would capture up to $5\times 10^{14}$ ($8\times 10^{17}$) dark matter particles per second---more than a factor of 100 higher.  Even when considering the larger surface area over which the resulting energy will be distributed, this represents considerably more power per square meter than is produced by dark matter inside of an Earth-like planet, although still far too little to meaningfully contribute to the temperatures required to maintain liquid water. 

In Fig.~\ref{local}, we show the capture rate (in energy-equivalent units) of dark matter particles onto an Earth-like planet in the local neighborhood of the Milky Way, as a function of the planet's mass. We explicitly carry out the calculation for masses of 1, 3 and 10 Earth masses, and interpolate/extrapolate (as shown as a dashed line) for other masses.  For this case, we assume a relative motion of the Solar System relative to the dark matter frame of 250 km/s, a velocity dispersion of 250 km/s, and a local density of 0.4 GeV/cm$^3$.

\section{Maintaining Liquid Water With Annihilating Dark Matter}
\label{3}

The surface temperature of a planet can be determined by equating the energy it absorbs to that it emits as an approximate blackbody:
\begin{equation}
P_{\rm star}+ P_{\rm geo} + P_{\rm DM} + ... = \sigma T^4_{\rm Pl} 4 \pi R_{\rm Pl}^2 \epsilon,
\end{equation}
where $\sigma$ is the Stefan-Boltzmann constant, $T_{\rm Pl}$ is the surface temperature of the planet, $R_{\rm Pl}$ is the radius of the planet, $\epsilon$ is the emissivity of the planet's atmosphere, and the subscripts ``star'', ``geo'', and ``DM'' correspond to the various contributions from the host star, planetary geological processes, and dark matter annihilations, etc.  In the case of energy from starlight, the incoming power is given by
\begin{equation}
P_{\rm star} = \sigma T^4_{\rm star} 4 \pi R_{\rm star}^2 \frac{\pi R^2_{\rm Pl}}{4 \pi D^2} (1-a),
\end{equation}
where $D$ is the distance between the star and planet, and $a$ is the planet's albedo.

As seen in Fig.~\ref{local}, dark matter annihilations are predicted to inject only on the order of $10^7$ to $10^9$ Watts or less into the surface region of an Earth-mass planet located in the local neighborhood of the Milky Way, even in the most optimistic models.  While the temperature that results from this energy input depends on the planet's emissivity, for Earth-like values ($\epsilon \approx 0.6$) this falls at least a factor of $10^7$ short of that needed to maintain liquid water in the absence of starlight.  A very different conclusion can be reached, however, if one considers planets in regions with much higher densities of slow moving dark matter.  As an example, we consider planets in the innermost tens of parsecs of a dwarf spheroidal galaxy.  Dwarf spheroidals are highly dense, and dark matter dominated systems.  \citet{list}, for example, describes seven dwarf spheroidals (Carina, Draco, Fornax, Leo I, Leo II, Sculptor, and Sextans) which have dark matter densities of 40-150 GeV/cm$^3$ and velocity dispersions of $\sim$10-20 km/s within their inner 10-20 parsecs. The lower velocities of these dark matter particles lead to a higher capture rate for two reasons: 1) They are more efficiently focused gravitationally toward the planet, and 2) They can become gravitationally bound to the planet after collisions in which they lose even a small amount of momentum. In such an environment, planets will be capable of capturing dark matter at much higher rates than are possible in our local neighborhood.

\begin{figure}[!ht]
\includegraphics[width=0.45\textwidth]{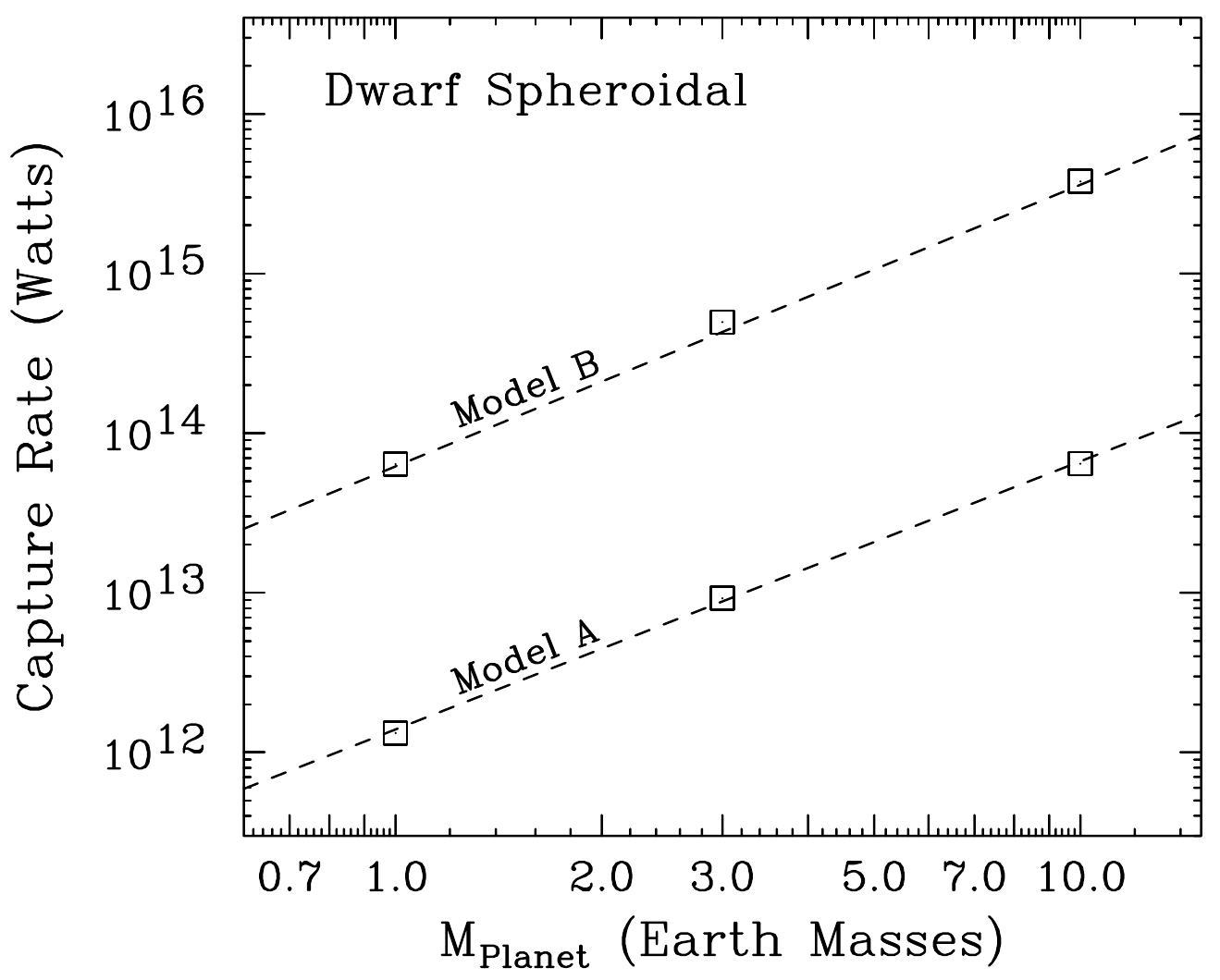}
\includegraphics[width=0.45\textwidth]{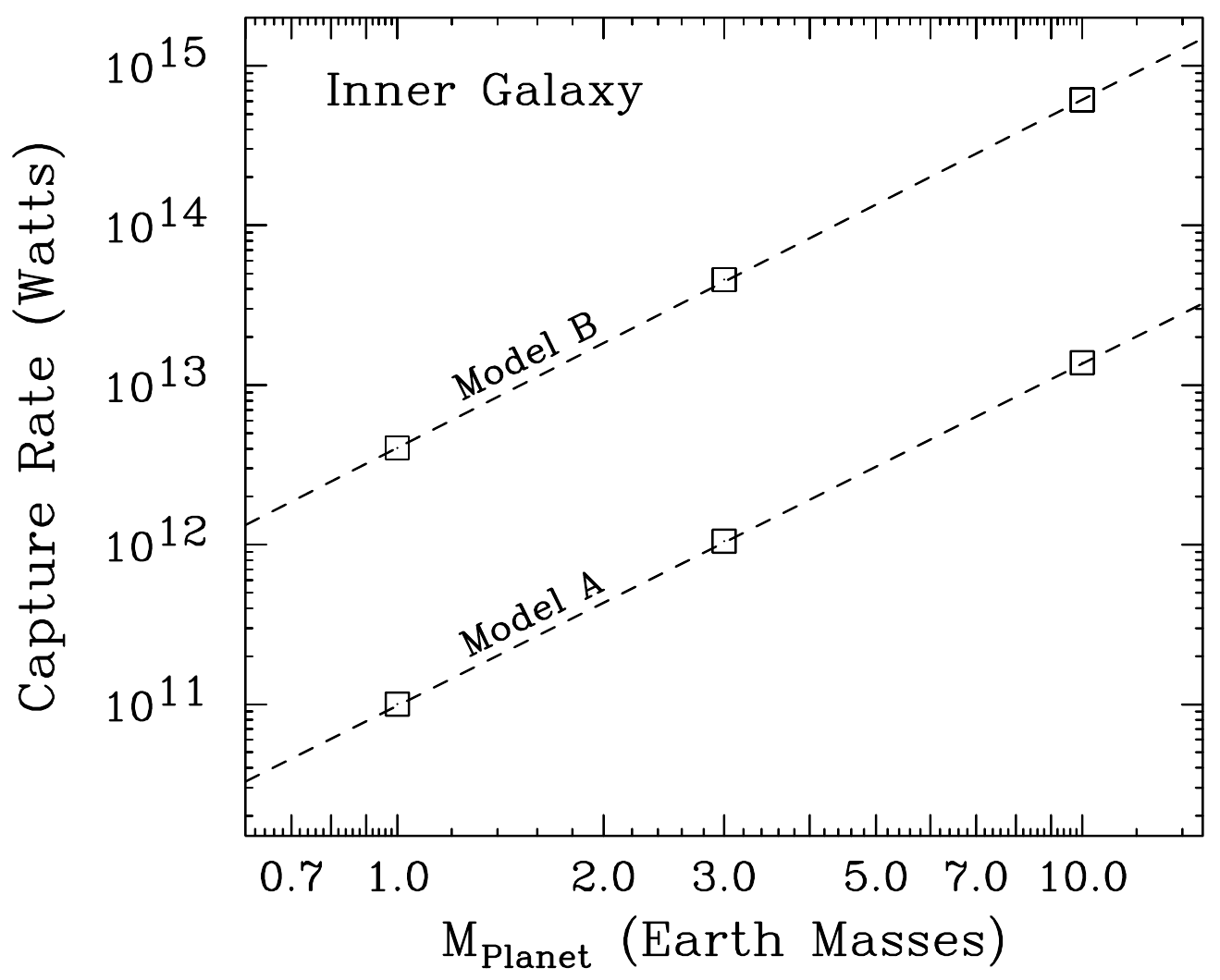}
\caption{The rate (in energy-equivalent units) at which dark matter particles are predicted to be captured by an Earth-like planet in the inner 10 parsecs of a dwarf spheroidal galaxy (top) and 10 parsecs from the center of the Milky Way (bottom), as a function of the planet's mass. We show results for two optimistic dark matter models. See text for more details. \label{capture}}
\end{figure}

In addition to planets in dwarf spheroidal galaxies, we also consider planets that are located very near the center of the Milky Way.  According to the frequently used Navarro-Frenk-White (NFW) profile \citep{nfw1,nfw2}, for example, the smooth component of the dark matter density in the Milky Way can be parametrized by
\begin{equation}
\rho_{\rm DM} \propto \frac{1}{r [1+(r/R_s)]^2},
\end{equation}
where $r$ is the distance to the Galactic Center, and $R_s\approx 20$ kpc is the scale radius.  Furthermore, the density in the innermost kiloparsecs of the Milky Way is generally expected to be higher than described by NFW due to baryonic adiabatic contraction~\citep{Prada:2004pi,Bertone:2005xv,Bertone:2005hw,Levine:2007zc}. To account for this, we alter the inner slope of the NFW halo profile such that $\rho_{\rm DM} \propto r^{-1.35}$ (instead of  $\rho_{\rm DM} \propto r^{-1.0}$) inside of 3 kpc of the Galactic Center.  We use for the velocity dispersion 50 km/s, consistent with measurements of the stellar velocity dispersion near the center of the galaxy \citep{Rangwala:2009}.

\begin{figure}[!ht]
\includegraphics[width=0.45\textwidth]{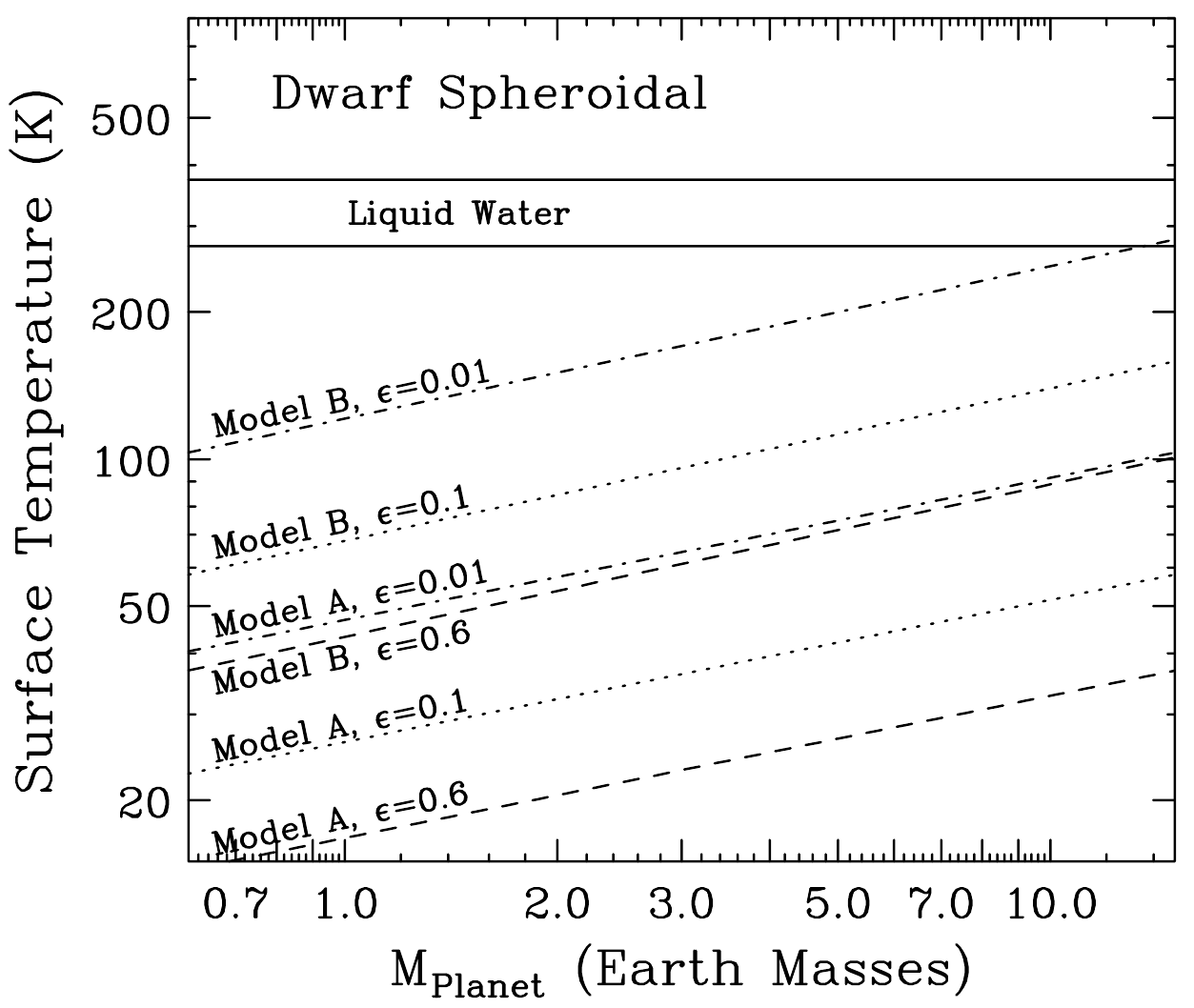} 
\includegraphics[width=0.45\textwidth]{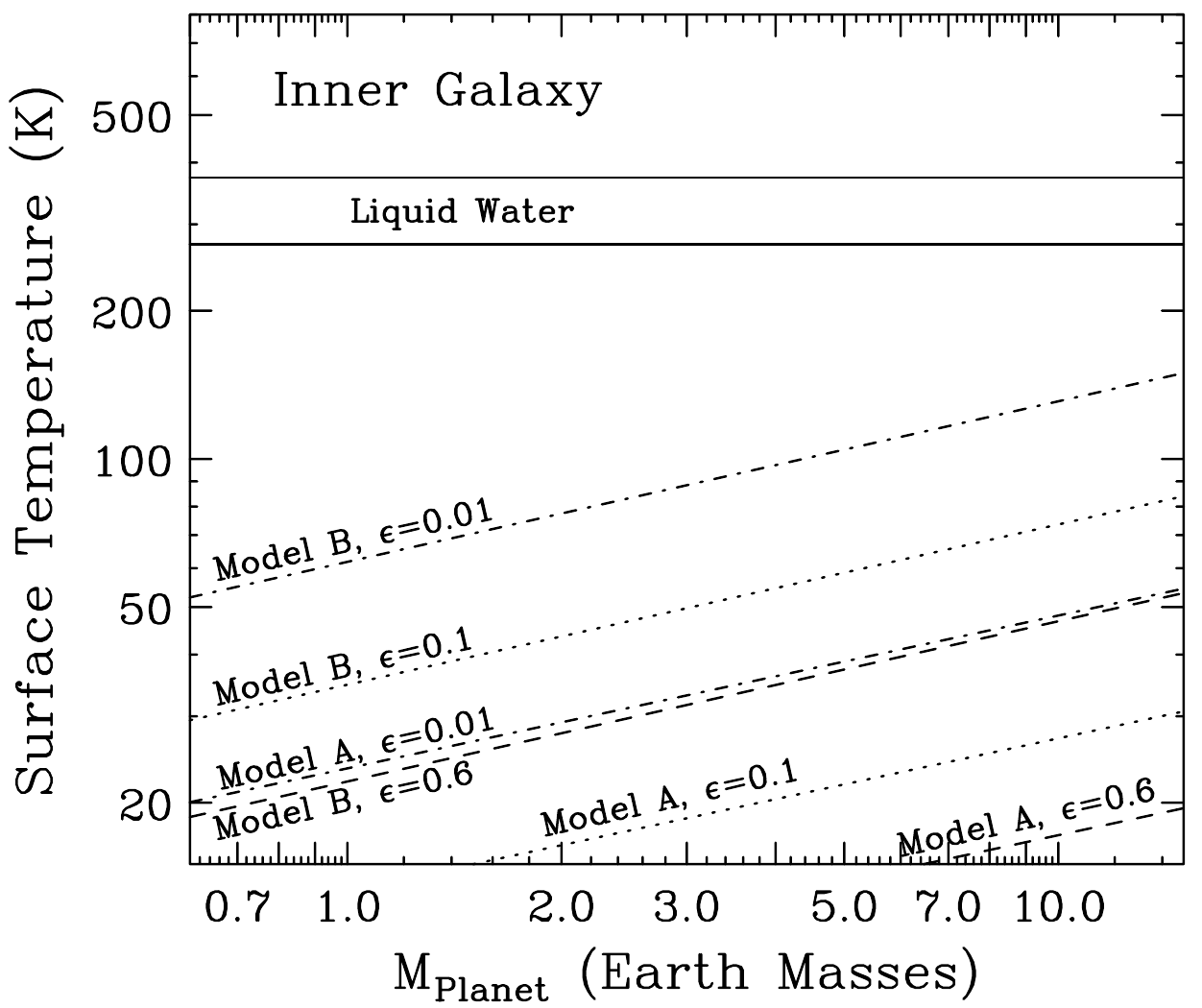} 
\caption{The surface temperature of an Earth-like-planet that is located in the inner 10 parsecs of a dwarf spheroidal galaxy (top) and 10 parsecs from the center of the Milky Way (bottom), as a function of the planet's mass. It is assumed that no significant starlight or other non-dark matter sources of energy contribute.  We show results for two optimistic dark matter models and for three values of the planets' emissivity, $\epsilon$.  In both scenarios there is an appreciable surface temperature from dark matter alone.  For the case of heavy planets with low emissivities, we find that surface temperatures can be high enough to sustain liquid water near the center of a dwarf spheroidal galaxy.  See text for more details.\label{temp}}
\end{figure}

In Fig.~\ref{capture}, we plot the capture rate of dark matter by an Earth-like planet, 10 parsecs from the center of a dwarf spheroidal galaxy ($\rho_{\rm DM}=150$ GeV/cm$^3$, $\sigma_v = 10$ km/s) and 10 parsecs from the center of the Milky Way ($\rho_{\rm DM}=825$ GeV/cm$^3$, $\sigma_v = 50$ km/s). Comparing these results to those shown in Fig.~\ref{local}, it is clear that planets in these environments are capable of accumulating dark matter at dramatically higher rates than are possible near our solar system.  If the dark matter can annihilate into standard model particles with a roughly weak-scale cross section, capture rates this large will easily lead to capture-annihilation equilibrium, {\it ie.}~the rate at which dark matter particles annihilate within a planet will be very nearly equal to the rate at which they are captured.

In Fig.~\ref{temp}, we show the surface temperature of the planets described in Fig.~\ref{capture}, in the absence of starlight or other non-dark matter sources of energy.  Here we assume that all of the annihilation products of the dark matter are absorbed by the surrounding planet.  If the dark matter annihilates significantly to neutrinos, the resulting surface temperature will be reduced accordingly.  We show results for three values of the emissivity, $\epsilon=$ 0.6, 0.1 and 0.01. A value of $\epsilon=$0.6 represents the value associated with the Earth's atmosphere.  More massive planets are generally expected to have denser atmospheres and thus lower emissivities.  With this in mind, we consider $\epsilon=$0.1--0.01 to be a conservative range for super-Earth planets.  None of the values we have considered represent an emissivity as low as that found on Venus ($\epsilon\approx 0.004$), for example.

Notably, from these calculations we see that planets in the innermost volume of the dwarf spheroidal galaxies, with masses of a few times that of the Earth or more could plausibly possess surface temperatures capable of sustaining liquid water if their atmospheres have sufficiently low emissivity.  
Other environments, such as near the core of the Large Magellenic Cloud (which is 10 times more massive than typical dwarf spheroidals), and other planets with particularly large masses of rocky material, such as a stripped-down version of the high density planet HD 149026 which has a rocky core of several tens of Earth-masses \citep{Sato:2005}, are also promising possibilities for high rates of planetary dark matter capture. 

To date, the planets observed closest to the Galactic Center have been those identified by the SWEEPS survey \citep{Sahu:2006}, although at a distance of $\sim$500 pc from the Galactic Center, we do not anticipate dark matter to provide a significant amount of energy to these planets---possibly comparable to the amount that the Earth receives from geothermal processes.

We point out that among the planets of the variety being considered here, the temperature will increase rapidly with depth from the surface. As a result, most of the planet will be molten, and only a thin (and possibly unstable) surface layer of solid material will be present. Any life to exist on such a planet will most likely be very different sort than that found on Earth.


%
\section{Discussion and Conclusions}
\label{discussion}

In this paper, we have calculated the capture rate of dark matter particles in Earth-like and super-Earth planets, and determined the resulting surface temperature of those planets that would result from dark matter annihilations. While planets in the local region of our galaxy receive only a negligible quantity of energy from dark matter annihilations, we find that planets in dwarf spheroidal galaxies (in particular) and in the innermost volume of the Milky Way could plausibly accumulate and annihilate enough dark matter to heat their surfaces to temperatures capable of sustaining liquid water provided that the emissivity of the atmosphere is sufficiently low.  The presence of liquid water could be maintained even in the absence of energy from starlight or other standard sources.

Although we expect ecologically relevant quantities of energy to be released through dark matter annihilations only within the interiors of planets that reside in very special environments (such as near the center of a dwarf spheroidal galaxy), and only in the case of dark matter models which feature large elastic scattering cross sections with nuclei (near the current upper limits), we expect that within such models planets will exist which derive enough heat from dark matter to almost indefinitely sustain surface temperatures sufficient to yield liquid water.  Even in the absence of starlight, such planets could plausibly contain life.  And, given their extremely long lifetimes, such planets may prove to be the ultimate bastion of life in our universe.

\bigskip

{\it Acknowledgements:}  The authors are supported by the US Department of Energy, including grant DE-FG02-95ER40896. DH is also supported by NASA grant NAG5-10842.%


\begin{thebibliography}{35}
\providecommand{\natexlab}[1]{#1}
\providecommand{\url}[1]{\texttt{#1}}
\expandafter\ifx\csname urlstyle\endcsname\relax
  \providecommand{\doi}[1]{doi: #1}\else
  \providecommand{\doi}{doi: \begingroup \urlstyle{rm}\Url}\fi

\bibitem[Aalseth et~al.(2010)]{cogent}
C.~E. Aalseth et~al.
\newblock \emph{arXiv:1002.4703}, 2010.

\bibitem[Abbot and Switzer(2011)]{Abbot:2011dz}
D.~S. Abbot and E.~R. Switzer.
\newblock \emph{arXiv:1102.1108}, 2011.

\bibitem[Ahmed et~al.(2010{\natexlab{a}})]{cdms}
Z.~Ahmed et~al.
\newblock \emph{Science}, 327:\penalty0 1619--1621, 2010{\natexlab{a}}.

\bibitem[Ahmed et~al.(2010{\natexlab{b}})]{cdmslow}
Z.~Ahmed et~al.
\newblock \emph{Phys.Rev.Lett.}, 2010{\natexlab{b}}.

\bibitem[Aprile et~al.(2010)]{xenon}
E.~Aprile et~al.
\newblock \emph{Phys.Rev.Lett.}, 105:\penalty0 131302, 2010.

\bibitem[Bernabei et~al.(2010)]{dama}
R.~Bernabei et~al.
\newblock \emph{Eur. Phys. J.}, C67:\penalty0 39--49, 2010.

\bibitem[Bertone and Merritt(2005{\natexlab{a}})]{Bertone:2005hw}
G.~Bertone and D.~Merritt.
\newblock \emph{Phys.Rev.}, D72:\penalty0 103502, 2005{\natexlab{a}}.

\bibitem[Bertone and Merritt(2005{\natexlab{b}})]{Bertone:2005xv}
G.~Bertone and D.~Merritt.
\newblock \emph{Mod.Phys.Lett.}, A20:\penalty0 1021, 2005{\natexlab{b}}.

\bibitem[Catena and Ullio(2010)]{Catena:2009mf}
R.~Catena and P.~Ullio.
\newblock \emph{JCAP}, 1008:\penalty0 004, 2010.

\bibitem[Duda et~al.(2007)Duda, Kemper, and Gondolo]{formfactor}
G.~Duda, A.~Kemper, and P.~Gondolo.
\newblock \emph{JCAP}, 0704:\penalty0 012, 2007.

\bibitem[Gould(1987)]{capture2}
A.~Gould.
\newblock \emph{Astrophys.J.}, 321:\penalty0 571, 1987.

\bibitem[Gould(1992)]{capture1}
A.~Gould.
\newblock \emph{Astrophys.J.}, 388:\penalty0 338, 1992.

\bibitem[Griest and Seckel(1987)]{capture3}
K.~Griest and D.~Seckel.
\newblock \emph{Nucl.Phys.}, B283:\penalty0 681, 1987.

\bibitem[Hooper et~al.(2010)Hooper, Collar, Hall, and McKinsey]{cogentdama}
D.~Hooper, J.I. Collar, J.~Hall, and D.~McKinsey.
\newblock \emph{Phys.Rev.}, D82:\penalty0 123509, 2010.

\bibitem[{Kasting} et~al.(1993){Kasting}, {Whitmire}, and
  {Reynolds}]{Kasting:1993}
J.~F. {Kasting}, D.~P. {Whitmire}, and R.~T. {Reynolds}.
\newblock \emph{Icarus}, 101:\penalty0 108--128, January 1993.

\bibitem[Krauss et~al.(1985)Krauss, Freese, Press, and Spergel]{Krauss:1985ks}
L.~M. Krauss, K.~Freese, W.~Press, and D.~Spergel.
\newblock \emph{Astrophys.J.}, 299:\penalty0 1001, 1985.

\bibitem[{Laughlin} et~al.(1997){Laughlin}, {Bodenheimer}, and
  {Adams}]{Laughlin:1997}
G.~{Laughlin}, P.~{Bodenheimer}, and F.~C. {Adams}.
\newblock \emph{\apj}, 482:\penalty0 420, June 1997.

\bibitem[{Laughlin} et~al.(2004){Laughlin}, {Bodenheimer}, and
  {Adams}]{Laughlin:2004}
G.~{Laughlin}, P.~{Bodenheimer}, and F.~C. {Adams}.
\newblock \emph{\apj L}, 612:\penalty0 L73--L76, September 2004.

\bibitem[Levine et~al.(2008)Levine, Gnedin, Hamilton, and
  Kravtsov]{Levine:2007zc}
R.~Levine, N.~Y. Gnedin, A.~J.~S. Hamilton, and A.~V. Kravtsov.
\newblock \emph{Astrophys.J.}, 678:\penalty0 154--167, 2008.

\bibitem[Lundberg and Edsjo(2004)]{Lundberg:2004dn}
J.~Lundberg and J.~Edsjo.
\newblock \emph{Phys.Rev.}, D69:\penalty0 123505, 2004.

\bibitem[Mack et~al.(2007)Mack, Beacom, and Bertone]{Mack:2007xj}
G.~D. Mack, J.~F. Beacom, and G.~Bertone.
\newblock \emph{Phys.Rev.}, D76:\penalty0 043523, 2007.

\bibitem[{Murray} and {Dermott}(2000)]{Murray:2000}
C.~D. {Murray} and S.~F. {Dermott}.
\newblock \emph{{Solar System Dynamics}}.
\newblock February 2000.

\bibitem[Navarro et~al.(1996)Navarro, Frenk, and White]{nfw1}
J.~F. Navarro, C.~S. Frenk, and S.~D.~M. White.
\newblock \emph{Astrophys.J.}, 462:\penalty0 563--575, 1996.

\bibitem[Navarro et~al.(1997)Navarro, Frenk, and White]{nfw2}
J.~F. Navarro, C.~S. Frenk, and S.~D.~M. White.
\newblock \emph{Astrophys.J.}, 490:\penalty0 493--508, 1997.

\bibitem[{Peale} and {Cassen}(1978)]{Peale:1978}
S.~J. {Peale} and P.~{Cassen}.
\newblock \emph{Icarus}, 36:\penalty0 245--269, November 1978.

\bibitem[{Pollack} et~al.(1993){Pollack}, {Hurter}, and
  {Johnson}]{Pollack:1993}
H.~N. {Pollack}, S.~J. {Hurter}, and J.~R. {Johnson}.
\newblock \emph{Reviews of Geophysics}, 1993.

\bibitem[Prada et~al.(2004)Prada, Klypin, Flix~Molina, Martinez, and
  Simonneau]{Prada:2004pi}
F.~Prada, A.~Klypin, J.~Flix~Molina, M.~Martinez, and E.~Simonneau.
\newblock \emph{Phys.Rev.Lett.}, 93:\penalty0 241301, 2004.

\bibitem[{Rangwala} et~al.(2009){Rangwala}, {Williams}, and
  {Stanek}]{Rangwala:2009}
N.~{Rangwala}, T.~B. {Williams}, and K.~Z. {Stanek}.
\newblock \emph{\apj}, 691:\penalty0 1387--1399, February 2009.

\bibitem[{Sahu} et~al.(2006)]{Sahu:2006}
K.~C. {Sahu} et~al.
\newblock \emph{\nat}, 443:\penalty0 534--540, October 2006.

\bibitem[{Sato} et~al.(2005)]{Sato:2005}
B.~{Sato} et~al.
\newblock \emph{\apj}, 633:\penalty0 465--473, November 2005.

\bibitem[Silk et~al.(1985)Silk, Olive, and Srednicki]{Silk:1985ax}
J.~Silk, K.~A. Olive, and M.~Srednicki.
\newblock \emph{Phys.Rev.Lett.}, 55:\penalty0 257--259, 1985.

\bibitem[{Stevenson}(1999)]{Stevenson:1999}
D.~J. {Stevenson}.
\newblock \emph{\nat}, 400:\penalty0 32, July 1999.

\bibitem[Valencia et~al.(2005)Valencia, O'Connell, and Sasselov]{densitypro1}
D.~Valencia, R.~J. O'Connell, and D.~D. Sasselov.
\newblock \emph{Icarus}, 2005.

\bibitem[Valencia et~al.(2007)Valencia, Sasselov, and O'Connell]{densitypro2}
D.~Valencia, D.~D. Sasselov, and R.~J. O'Connell.
\newblock \emph{Astrophys.J.}, 665:\penalty0 1413--1420, 2007.

\bibitem[Walker et~al.(2007)Walker, Mateo, Olszewski, Gnedin, Wang,
  et~al.]{list}
M.~G. Walker, M.~Mateo, E.~W. Olszewski, O.~Y. Gnedin, X.~Wang, et~al.
\newblock \emph{arXiv:0708.0010}, 2007.

\end{thebibliography}

\end{document}